\documentclass[12pt]{article}
\usepackage{graphicx}
\usepackage{epstopdf}
\usepackage{caption}
\usepackage{subcaption}
\usepackage{subfig}

\usepackage{epsfig,amsmath,amssymb,verbatim,mathrsfs}

\def\beq{\begin{eqnarray}}
\def\eeq{\end{eqnarray}}
\def\bea{\begin{eqnarray}}
\def\eea{\end{eqnarray}}

\newcommand{\dm}{\text{\tiny DM}}
\newcommand{\f}{\mathcal{F}}

\renewcommand{\thefootnote}{\roman{footnote}}

\begin{document}

\setlength{\baselineskip}{0.2in}


\begin{titlepage}
\noindent
\vspace{0.2cm}

\begin{center}
  \begin{Large}
    \begin{bf}
Scale-Invariant Models with One-Loop Neutrino Mass and Dark Matter Candidates\\

     \end{bf}
  \end{Large}
\end{center}

\vspace{0.2cm}

\begin{center}

\begin{bf}

{Amine~Ahriche,$^{1,2,}$\footnote{aahriche@ictp.it} Adrian~Manning,$^{3,}$\footnote{adrian.manning@sydney.edu.au} 
Kristian~L.~McDonald$^{3,}$\footnote{kristian.mcdonald@sydney.edu.au} and
Salah~Nasri$^{4,5,}$\footnote{snasri@uaeu.ac.ae}}\\
\end{bf}
\vspace{0.5cm}
 \begin{it}
$^1$ Department of Physics, University of Jijel, PB 98 Ouled Aissa, DZ-18000 Jijel, Algeria\\
\vspace{0.1cm}
$^2$ The Abdus Salam International Centre for Theoretical Physics, Strada Costiera 11, I-34014, Trieste, Italy\\
\vspace{0.1cm}
$^3$ ARC Centre of Excellence for Particle Physics at the Terascale,\\
School of Physics, The University of Sydney, NSW 2006, Australia\\
\vspace{0.1cm}
$^4$ Physics Department, UAE University, POB 17551, Al Ain, United Arab Emirates\\
\vspace{0.1cm}
$^5$
Laboratoire de Physique Theorique, Es-Senia, University, DZ-31000, Oran, Algeria  
 \vspace{0.3cm}
\end{it}
\vspace{0.5cm}

\end{center}


\begin{abstract}

We construct a list of minimal scale-invariant models at the TeV scale that generate one-loop neutrino mass and give viable dark matter candidates. The models generically contain a singlet scalar and a $Z_2$-odd sector comprised of singlet, doublet and/or triplet SU(2) multiplets. The dark matter may reside in a single multiplet or arise as an admixture of several multiplets. We find fifteen independent models, for which the dark matter is a viable candidate and neutrino mass results from a diagram with just one of the irreducible scale-invariant one-loop topologies. Further ``non-pure" cases give hybrid one-/two-loop masses. All models predict new TeV scale physics, including a singlet scalar that generically mixes with the Higgs boson.

\end{abstract}

\vspace{1cm}

\end{titlepage}
\renewcommand{\thefootnote}{\arabic{footnote}}
\setcounter{footnote}{0}
\setcounter{page}{1}


\vfill\eject


\section{Introduction\label{sec:introduction}}

A number of the remaining puzzles in particle physics relate to our incomplete understanding of the mechanisms of mass in the universe. Such puzzles include the nature of dark matter (DM), the mechanism of neutrino mass, and the origin of the $\mathcal{O}(100)$~GeV Higgs mass-parameter that determines the weak scale in the standard model (SM). With regards to the last issue, a number of recent works have studied extensions of the SM that possess a scale-invariance (SI) symmetry, such that the Higgs mass arises as a quantum effect via radiative (Coleman-Weinberg~\cite{Coleman:1973jx}) symmetry breaking~\cite{Hempfling:1996ht}. These models can have interesting phenomenology and generally predict new particles at or around the TeV scale (for recent analyses see e.g.~Ref.~\cite{Farzinnia:2015uma}).

Adopting a SI symmetry modifies the way in which candidate solutions to outstanding problems are implemented and opens up new approaches. For example, the origin of neutrino mass can find interesting explanations within SI models~\cite{Foot:2007ay,Iso:2009ss,Ahriche:2016cio,Lee:2012jn} (for detailed discussion see Ref.~\cite{Lindner:2014oea}). Among the available possibilities, it is perhaps an obvious marriage to employ a radiative mechanism for neutrino mass  within the SI context, giving a common radiative origin for both neutrino mass and the weak scale. Such models typically require beyond-SM fields, to  both trigger  electroweak symmetry breaking and allow radiative neutrino mass. An interesting approach is to consider extensions of the SM where the new multiplets permitting  radiative symmetry breaking also give rise to neutrino mass and DM.

Motivated by these considerations, in this work we compile a list of minimal SI models that generate one-loop radiative neutrino mass while giving a  viable DM candidate. In the process we catalogue the minimal irreducible SI one-loop topologies for neutrino mass (defined in the text). We focus primarily on  models where neutrino mass results from \emph{just one} of the minimal irreducible SI one-loop topologies; models generating topologically distinct one-loop diagrams can be considered as generalized versions of the model realizing the simpler topology. We find fifteen independent models realizing neutrino mass via a single SI one-loop topology, the simplest of which is the SI scotogenic model (recently studied in detail~\cite{Ahriche:2016cio}). The fifteen independent models are listed in Tables~\ref{table:general_SIMa},~\ref{table:SIT1-2} and \ref{table:SIT1-3} in the text. Further models employ a one-loop topology that also allows a two-loop diagram with lower mass-dimension, meaning they are not  ``pure" one-loop models. This differs from the non-SI case, where the analogous topology gives pure one-loop models. In addition, the SI models generically contain a singlet scalar that participates in electroweak symmetry breaking and births the requisite lepton number symmetry breaking. This field mixes with the SM Higgs. The models also contain a TeV scale sector comprised of singlets, doublets and/or triplets, which participate in the neutrino mass diagram and include a DM candidate. 

The structure of this paper is as follows. In Section~\ref{sec:prelim} we provide general preliminaries for our analysis. The minimal irreducible SI one-loop topologies for neutrino mass  are described in Section~\ref{sec:variant_models}, where corresponding lists of viable models with DM candidates are presented. Conclusions are drawn in Section~\ref{sec:conc}. Before proceeding we note that earlier authors have studied relationships between neutrino mass and DM; see e.g.~Refs.~\cite{Krauss:2002px,Ma:2006km,Ma:2008cu,Aoki:2011yk,Aoki:2013gzs} and Refs.\cite{Law:2013saa,Brdar:2013iea,Ng:2014pqa,Culjak:2015qja,Restrepo:2013aga}.


\section{General Preliminaries\label{sec:prelim}}

\begin{figure}[ttt]
\begin{center}
        \includegraphics[width = 0.50\textwidth]{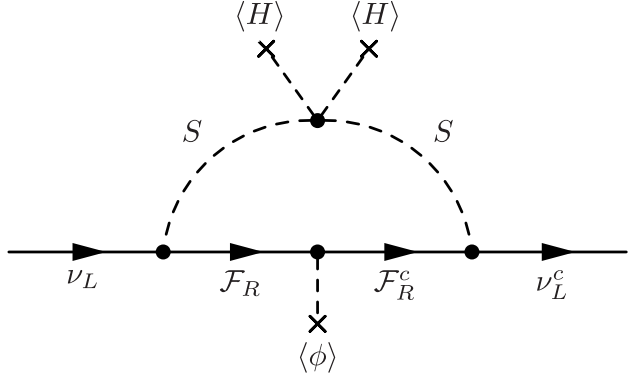}
\end{center}
\caption{One-loop diagram for neutrino mass in the scale-invariant scotogenic model.}\label{fig:SI_scotogenic}
\end{figure}

A recent paper performed a detailed analysis of the minimal SI scotogenic model~\cite{Ahriche:2016cio}, demonstrating the existence of viable parameter space, consistent with both flavor and direct-detection constraints. The model is implemented by extending the SM to include three generations of gauge-singlet fermions, $\f_{iR}\sim(1,1,0)$, where $i=1,\,2,\,3,$ labels generations, a second SM-like scalar doublet, $S\sim(1,2,1)$, and a singlet  scalar $\phi\sim(1,1,0)$. In addition to the SI symmetry, a $Z_2$ symmetry, with action $\{\f_R,\, S\}\rightarrow - \ \{\f_R,\, S\}$, is imposed, with  the scalar $\phi$ and the SM fields  transforming trivially under this symmetry. The lightest $Z_2$-odd particle is a stable DM candidate; this should be taken as either the lightest singlet fermion $\f_{1}$ or a neutral component of the doublet $S$. However, viable symmetry breaking requires one of the beyond-SM scalars to be the heaviest exotic multiplet, making fermionic DM more likely. The scalar $\phi$ plays the dual roles of sourcing lepton number symmetry breaking, to allow neutrino masses, and triggering electroweak symmetry breaking. Neutrinos acquire mass via the one-loop diagram shown in Figure~\ref{fig:SI_scotogenic} (here $H\sim(1,2,1)$ denotes the SM scalar doublet). The SI scotogenic model was also mentioned in Refs.~\cite{Lee:2012jn,Lindner:2014oea}.

The SI scotogenic model belongs to a larger family of SI models with one-loop neutrino mass and DM. In this work we catalogue the minimal  implementation of these related  models. There are four distinct classes of models, categorized by the topology of the corresponding one-loop diagram. The representative mass diagrams for these classes of models comprise the set of minimal SI one-loop topologies for neutrino mass. One class contains the SI scotogenic model and its related variants, while the others employ distinct one-loop diagrams. 

Before describing the variant models, we outline some of their general features. The minimal SI implementation of all models includes a singlet scalar $\phi$, which, similar to the SI scotogenic model, continues to play a role in electroweak symmetry breaking and in allowing lepton number symmetry violation. In addition, the models employ a set of beyond-SM fermions $\f$ and scalars $S$, which are odd under a $Z_2$ symmetry, and thus contain a DM candidate. The $Z_2$-odd scalars must have a vanishing vacuum value (VEV), $\langle S\rangle=0$, to preserve the $Z_2$ symmetry and ensure DM longevity. The ground state has $\langle H\rangle\ne0$ and $\langle \phi\rangle\ne0$, with both VEVs playing a role in the neutrino mass diagram. The analysis of the full one-loop potential is somewhat involved (see e.g.~\cite{Ahriche:2016cio}), though in general, absent parameter hierarchies, one expects $\langle \phi\rangle\sim \langle H\rangle$, as the VEVs are related via dimensionless couplings. The SM Higgs and $\phi$ mix, so the spectrum contains two physical $Z_2$-even scalars $h_{1,2}$, one of which has mass $M_{h_1}\simeq 125$~GeV, and is the SM-like scalar, while the other is the pseudo-Goldstone boson (or dilaton) associated with broken SI symmetry. The latter acquires mass at the one-loop level, and the demand that $M_{h_2}>0$ exhibits a constraint on the spectrum. The dilaton mass is well-approximated by~\cite{Gildener:1976ih}
\begin{equation}
M_{h_{2}}^{2}\simeq \frac{1}{8\pi ^{2}(\langle \phi\rangle^{2}+\langle H\rangle^{2})}\left\{
M_{h_{1}}^{4}+6M_{W}^{4}+3M_{Z}^{4}-12M_{t}^{4}+\sum_S n_S M_{S}^{4}-\sum_{\f}n_\f M_{\f}^{4}\right\} ,  \label{eq:dilaton_mass}
\end{equation}
where the sum is over all beyond-SM particles (except the dilaton) and $n_{S,\f}$ are multiplicity factors. One of the scalars $S$ must be the heaviest exotic in the spectrum to overcome negative loop-corrections from both the top quark and the exotic fermions $\f$. In models with a single exotic scalar $S$, the DM should be fermionic.

The mixing between $H$ and $\phi$ has important consequences. Any field that couples to $\phi$ inherits a coupling to the SM sector via the mixing with $H$. This includes the DM, which typically acquires its mass via a coupling to $\phi$, due to the absence of bare mass terms in the SI theory. This can give additional annihilation channels for the DM (into $h_1 h_2$ final states), and additional couplings between the DM and quarks, of immediate relevance for direct-detection experiments. The mixing between $\phi$ and $H$ is typically controlled by the ratio of VEVs, and cannot be made arbitrarily small without introducing parameter hierarchies. Thus, the models are typically subject to stringent direct-detection constraints, as analyzed in detail for the SI scotogenic model in Ref.~\cite{Ahriche:2016cio} and  the SI KNT model in Ref.~\cite{Ahriche:2015loa}. Consequently, ongoing and future direct-detection experiments will provide useful information on the  models.

The DM should typically be the neutral component of a hypercharge-less multiplet, to avoid exclusion via direct-detection constraints (due to $Z$ boson exchange). This demand can be alleviated if e.g.~the CP odd and CP even components of a complex DM candidate can be split.  In this regard, models in which the DM has nonzero hypercharge are generally disfavored by direct-detection experiments, with the following exceptions: models with an SM-like scalar, $S\sim(1,2,-1)$, can be viable due to the splitting induced by the term $(SH)^2$, which is always allowed by both the SI and $Z_2$ symmetries; models with a complex scalar triplet can be allowed, provided the triplet mixes with either a real scalar or a doublet scalar, as both can induce the requisite splitting of the neutral component; models with a complex fermion can be allowed, if the neutral component of the complex fermion mixes with a  real fermion, to provide the splitting (whereas models with two complex fermions that mix to give $U(1)_Y$-charged Dirac DM are generally not viable). 

The models generically contain a vertex of the form $L S \f$ for an exotic scalar $S$ and fermion $\f$. Such vertices can induce lepton flavor violating (LFV) effects, like $\mu\rightarrow e+\gamma$. The existence of charged scalars and new fermions that couple to the Higgs (through the mixing of $H$ and $\phi$) also affects the electroweak precision observables and the Higgs decay width $\Gamma(h\rightarrow \gamma\gamma)$. Thus,  LFV searches, electroweak precision measurements and  $h\rightarrow \gamma\gamma$ decays all provide useful ways to test the models and/or constrain the parameter space. The severity of the constraints are subject to model dependencies, though, in general the  analyses are similar to those detailed  in Ref.~\cite{Ahriche:2016cio} for the  SI scotogenic model.

In categorizing the models, in what follows, we exclude cases with complex DM unless the requisite splitting is automatically achieved by the minimal particle content needed for the neutrino mass diagram. We also exclude cases where the particle content required to generate a given one-loop neutrino mass diagram includes a real fermion and a scalar doublet $S\sim(1,2,1)$. Such models also generate the one-loop diagram from the SI scotogenic model (or the related triplet variant) and can be considered as generalized versions of the model defined by the simpler subset of particles. Furthermore, we restrict our attention to new multiplets no larger than the adjoint representation, and fields whose electric charge is no higher than doubly charged (in units of the proton charge). We now turn to our classification, beginning with the models most closely related to the SI scotogenic model.

\section{Minimal Scale-Invariant One-Loop Models\label{sec:variant_models}}

\subsection{Scale-Invariant Type T3 Models}
\begin{figure}[ttt]
\begin{center}
        \includegraphics[width = 0.50\textwidth]{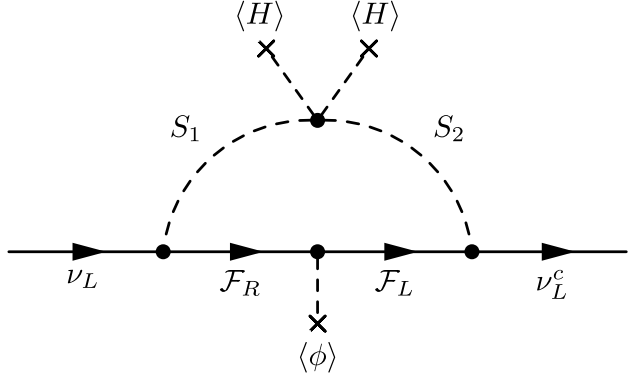}
\end{center}
\caption{One-loop diagram for neutrino mass via the scale-invariant type T3 topology. The scale-invariant scotogenic model gives the simplest realization.}\label{fig:general_SI_1loop}
\end{figure}

\begin{table}
\centering
\begin{tabular}{|c|c|c|c|c|c|}\hline
& & &  &&Related\\
\ \ SI Type T3 Models\ \ &
$\mathcal{F}$  & $S_1$&$S_2$&Dark Matter& Non-SI\\
& & &  && Model\\
\hline
SI Scotogenic& $(1,1,0)$ &$(1,2,1)$&$-$&Singlet Fermion&\cite{Ma:2006km}\\ 
\hline
 SI Triplet Scotogenic& $(1,3,0)$ &$(1,2,1)$ &$-$&Triplet Fermion&\cite{Ma:2008cu}\\ 
\hline
$(a)$& $\ \ (1,1,-2)\ \ $ &$\ \ (1,2,1)\ \ $&\ \ $(1,2,3)$\ \ & Doublet Scalar& \cite{Aoki:2011yk}\\ 
\hline
$(b)$& $\ \ (1,3,-2)\ \ $ &$\ \ (1,2,1)\ \ $&\ \ $(1,2,3)$\ \ & Doublet Scalar&\cite{Law:2013saa}\\ 
\hline
$(c)$& $(1,2,-1)$ &$(1,1,0)$&\ $(1,3,2)$\ & Singlet-Triplet Scalar&\cite{Law:2013saa}\\ 
\hline
$(d)$& $(1,2,-1)$ &$(1,3,0)$&$(1,1,2)$&  Triplet Scalar &\cite{Law:2013saa,Brdar:2013iea}\\ 
\hline
$(e)$& $(1,2,-1)$ &$(1,3,0)$ &$(1,3,2)$& Triplet Scalar&\cite{Law:2013saa}\\ 
\hline
\end{tabular}
\caption{Models with dark matter candidates and one-loop neutrino mass  via the  scale-invariant Type T3 topology (see Figure~\ref{fig:general_SI_1loop}). All  models contain  a singlet scalar $\phi\sim(1,1,0)$, and a discrete symmetry $\{\f,S_{1,2}\}\rightarrow-\{\f,S_{1,2}\}$.  Here, $\f$ is a vector-like (chiral) fermion for $Y\ne0$ ($Y=0$), and $S_1=S_2$ for models with chiral fermions.\label{table:general_SIMa} }
\end{table}


There exist generalizations of the scotogenic model that generate neutrino mass via a one-loop diagram with the same topology~\cite{Law:2013saa}. One can consider minimal  SI implementations of these variant models. The general one-loop diagram for neutrino mass in such models is shown in Figure~\ref{fig:general_SI_1loop}, with model-dependent quantum numbers for the  intermediate fields.  Precluding cases where the DM is already excluded, the  particle content for viable implementations is given in  Table~\ref{table:general_SIMa}. The fermion $\f$ is taken as vector-like in cases where it is complex-valued, and all cases utilize a $Z_2$ symmetry with action $\{\f,S_1,S_2\}\rightarrow-\{\f,S_1,S_2\}$.  Note that the diagram for neutrino mass in the SI case has mass-dimension six. 

One observes  that, in addition to the triplet variant of the SI scotogenic model, there are models with singlet, doublet or triplet scalar DM candidates. Models with doublet scalar DM ($S_1$) utilise the mixing term $(H^\dagger S_1)^2$ to split the neutral components of $S_1$, and allow consistency with direct-detection constraints. If the DM is  a real singlet or triplet scalar, it does not couple to the $Z$ boson and direct-detection constraints can typically be evaded. For complex triplet DM, $S_2\sim(1,3,2)$, the DM would  usually be excluded. However, the models generate mixing between the complex triplet $S_2$ and the real scalar $S_1$, to split the neutral components of $S_2$~\cite{Law:2013saa}. Thus, one cannot rule out these cases, a~priori.

The table shows that a number of the variant models have DM candidates that belong to an $SU(2)_L$ triplet. Due to the nontrivial $SU(2)_L$ gauge interactions, such DM candidates require  heavier  masses of  $M_\dm\approx2-3$~TeV. Although it is possible to generate these larger masses in  SI models, a degree of tuning is required to retain a Higgs mass of 125~GeV with electroweak triplets above the TeV scale. This is understood as follows. The heavier exotics do not decouple from the SM sector in the limit that one (or more) Yukawa and/or scalar couplings vanish~\cite{Foot:2013hna}. Thus, one cannot sequester the exotics from the SM to shield the Higgs mass. Consequently  higher-order corrections involving the heavy fields give naturalness constraints that typically require $M_{heavy}<$~TeV, in tension with the demand of $M_\dm=\mathrm{few}~$TeV. Of course,  the amount of fine-tuning is not severe for $M_\dm=\mathrm{few}~$TeV and such models may be of phenomenological interest.

Alternatively, models $(a)$, $(b)$ and $(c)$ contain DM candidates whose mass need not exceed the TeV scale. Models   $(a)$ and $(a)$  give inert-doublet DM, which does not require heavy fields with non-trivial electroweak quantum numbers. Similarly, model $(c)$ admits inert-singlet DM. These models contain Dirac fermions, as evidenced by the nonzero hypercharge values in Table~\ref{table:general_SIMa}. The DM must be a scalar  and  the mass ordering should be
\bea
M_\dm\ <\ M_{\f_i} \ <\ M_{S>},
\eea
where $M_{S>}$ denotes the (approximately) common mass for the scalar multiplet that does not contain the DM. The heaviest exotic must  be a scalar to ensure $M_{h_2}>0$. These models are not obviously excluded and may deserve further study.

\begin{figure}[ttt]
\begin{center}
        \includegraphics[width = 0.50\textwidth]{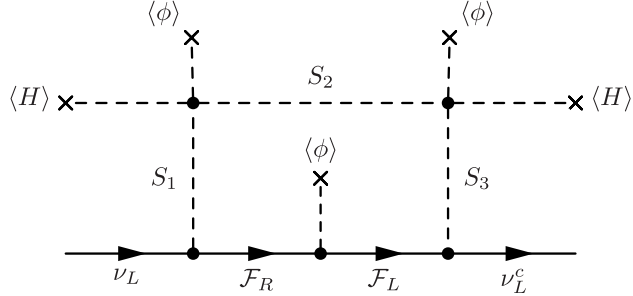}
\end{center}
\caption{One-loop diagram for neutrino mass via the scale-invariant type T1-i topology..}\label{fig:SI_T1-1}
\end{figure}

The set of models with particle content in Table~\ref{table:general_SIMa}, which achieve one-loop neutrino mass via Figure~\ref{fig:general_SI_1loop}, could be called the SI type T3 one-loop models as they are the minimal SI implementation of the type T3 one-loop topology for neutrino mass (we refer to the labeling scheme of Ref.~\cite{Bonnet:2012kz}). This set includes the SI scotogenic model and its triplet variant, in addition to five models with Dirac fermions. These SI models are related to non-SI models that exist in the literature, as listed in the final column  of Table~\ref{table:general_SIMa}.  In addition to these minimal SI T3 models, one can also consider minimal SI implementations of the alternative irreducible one-loop topologies, as we now discuss. 
\subsection{Scale-Invariant Type T1-i Models}
First we consider the minimal SI implementation of the type T1-i models, for which neutrino mass arises via the one-loop  diagram in Figure~\ref{fig:SI_T1-1}. Comparison of Figures~\ref{fig:general_SI_1loop} and~\ref{fig:SI_T1-1} reveals that the loop diagrams have some similarities. In particular, the SI T1-i  diagram in Figure~\ref{fig:SI_T1-1} can be generated by ``opening up" the top vertex in Figure~\ref{fig:general_SI_1loop}, attaching two external $\phi$ VEVs, and including an extra internal scalar particle (labeled as $S_2$ in Figure~\ref{fig:SI_T1-1}). Thus, the minimal particle content required to generate an SI T1-i model always contains a subset of particles that generates an SI T3 diagram. Consequently one can consider the type T1-i models as generalized versions of the T3 models; for an SI T1-i model, neutrino mass  always receives contributions from both the T3 and T1-i diagrams: $m_\nu = m_{T3}+m_{T1-i}$. Nonetheless, the T1-i models may still be of phenomenological interest, as one can always select parameter space in which the T1-i diagram is dominant, namely by taking the coupling for the $S_1S_2 H^2$ vertex in Figure~\ref{fig:general_SI_1loop} to be very small, such that $m_{T3}\ll m_{T1-i}$, which gives $m_\nu=m_{T3}+m_{T1-i}\approx m_{T1-i}$.  Given the relationship between the T3 and T1-i models, we do not list explicit quantum numbers for candidate models - these can be obtained by selecting a T3 model from Table~\ref{table:general_SIMa} and finding the quantum numbers for the requisite additional multiplet labeled as $S_2$ in Figure~\ref{fig:SI_T1-1}. We note that the SI T1-i diagram has mass-dimension eight; we discuss this matter below.

\begin{figure}[ttt]
\begin{center}
        \includegraphics[width = 0.50\textwidth]{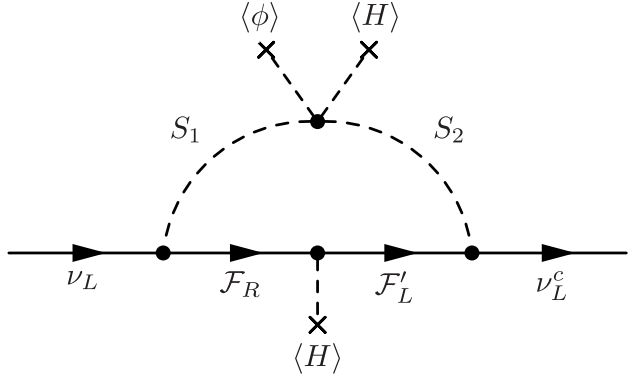}
\end{center}
\caption{One-loop diagram for neutrino mass via the scale-invariant type T1-ii topology.}\label{fig:SI_T1-2}
\end{figure}

\begin{table}
\centering
\begin{tabular}{|c|c|c|c|c|c|}\hline
& & &  & &\\
\ \ SI Type T1-ii Models\ \ &
$\f$  & $\f'$&$S_1$&$S_2$&Dark Matter\\
& & &  & &\\
\hline
(a)& $(1,1,-2)$ &$(1,2,-1)$&$(1,2,1)$&$(1,1,2)$& Doublet Scalar \\ 
\hline
(b)& $(1,1,-2)$ &$(1,2,-1)$&$(1,2,1)$&$(1,3,2)$& Doublet-Triplet Scalar\\ 
\hline
(c)& $(1,2,1)$ &$(1,3,2)$&$(1,1,-2)$&$(1,2,-1)$& Doublet Scalar \\ 
\hline
(d)& $(1,2,1)$ &$(1,3,2)$&$(1,3,-2)$&$(1,2,-1)$& Doublet Scalar\\ 
\hline
\end{tabular}
\caption{ Scale-invariant models with dark matter candidates and one-loop neutrino mass  via the type T1-ii topology (Figure~\ref{fig:SI_T1-2}). All  models contain  a singlet scalar $\phi\sim(1,1,0)$, and a discrete symmetry $\{\f,\f',S_1,S_2\}\rightarrow-\{\f,\f',S_1,S_2\}$, where $\f$ and $\f'$ are vector-like fermions.\label{table:SIT1-2} }
\end{table}


\subsection{Scale-Invariant Type T1-ii Models}

Next, we consider models that give neutrino mass via the SI T1-ii one-loop topology, as shown in  Figure~\ref{fig:SI_T1-2}. We find four independent models with this topology, as listed in Table~\ref{table:SIT1-2}. The models have vector-like fermions $\f$ and $\f'$, and two beyond-SM scalar multiplets $S_{1,2}$, all odd under a $Z_2$ symmetry. The fermions need not be vector-like to generate the one-loop diagram but using vector-like fermions is the simplest way to avoid anomalies. The scalar $S_2$ should be the heaviest exotic, while the DM multiplet $S_1\sim(1,2,1)$ is the lightest. Additional variant models which contain either fermion triplets or singlets with vanishing hypercharge are not considered, as they automatically contain a doublet scalar and consequently also generate the SI type-T3  diagram. All four models in Table~\ref{table:SIT1-2} give scalar DM and contain an SM-like doublet scalar, which can be the DM. In addition, model T1-ii$(b)$ can give doublet-triplet DM, while models T1-ii$(c)$ and T1-ii$(d)$ can give singlet or singlet-doublet DM. We note that if one does not seek to include a  DM candidate, such that SM fields may propagate in the loop diagram, the SI T1-ii topology describes the SI  implementation of the Zee model~\cite{Foot:2007ay,Lindner:2014oea}. We also note that non-SI versions of the models in Table~\ref{table:SIT1-2} appeared in Ref.~\cite{Restrepo:2013aga}.

\begin{figure}[ttt]
\begin{center}
        \includegraphics[width = 0.50\textwidth]{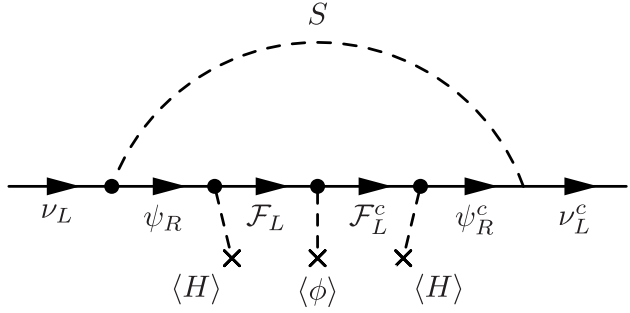}
\end{center}
\caption{One-loop diagram for neutrino mass via the scale-invariant type T1-iii topology.}\label{fig:SI_T1-3}
\end{figure}

\begin{table}
\centering
\begin{tabular}{|c|c|c|c|c|}\hline
& & &  &\\
\ \ SI Type T1-iii Models\ \ &
$\mathcal{F}$  & $\psi$&$S$&Dark Matter\\
& & &  &\\
\hline
(a)& $(1,1,0)$ &$(1,2,-1)$&$(1,1,0)$&Singlet-Doublet Fermion \\ 
\hline
(b)& $(1,1,0)$ &$(1,2,-1)$&$(1,3,0)$&Singlet-Doublet Fermion\\ 
\hline
(c)& $(1,3,0)$ &$(1,2,-1)$&$(1,3,0)$&Triplet-Doublet Fermion\\ 
\hline
(d)& $(1,3,0)$ &$(1,2,-1)$&$(1,1,0)$&Triplet-Doublet Fermion\\ 
\hline
\end{tabular}
\caption{ Scale-invariant models with dark matter candidates and one-loop neutrino masses  via the type T1-iii topology (Figure~\ref{fig:SI_T1-3}). All  models contain  a singlet scalar $\phi\sim(1,1,0)$, and a discrete symmetry $\{\f,\psi,S\}\rightarrow-\{\f,\psi,S\}$, where $\psi$ is a vector-like fermion.\label{table:SIT1-3} }
\end{table}

\subsection{Scale-Invariant Type T1-iii Models}
Finally we consider models with the SI type T1-iii one-loop topology, as shown in Figure~\ref{fig:SI_T1-3}. In this case the mass diagram has mass-dimension six. In addition to $\phi$, these models include a single beyond-SM scalar $S$. Even if this multiplet contains a DM candidate, one must restrict attention to parameter space in which $S$ is the heaviest exotic, in order to dominate the fermionic contributions to the effective potential and ensure a non-negative dilaton mass. This preferences cases with fermionic DM. Including this consideration, we find four models, listed as T1-iii$(a)$ through T1-iii$(d)$ in Table~\ref{table:SIT1-3}. These contain two real fermions $\f$ and $\psi$, both odd under the $Z_2$ symmetry (along with $S$). All have real fermionic DM, with either triplet or singlet fermions possible,  which, in general, mix with the doublet fermion.   There may appear to be additional models to those listed in Table~\ref{table:SIT1-3}. However, these contain an SM-like  doublet scalar $S\sim(1,2,1)$ and a fermion transforming as either a singlet $(1,1,0)$, or a triplet $(1,3,0)$. Thus, these models include the same particle content as the first two models in Table~\ref{table:general_SIMa}, and automatically generate a diagram with the SI type T3 topology. 

One can also consider models with the real fermion $\f$ in Figure~\ref{fig:SI_T1-3} replaced by a complex fermion, and $\psi_R^c$ replaced by an independent field $\psi_L'$. However, we find no viable realizations in this case. There are candidate theories containing neutral fields, which could give DM,  though in all instances the DM is excluded  by direct-detection constraints, or the model also gives a T3 diagram. Note that non-SI versions of the models in Table~\ref{table:SIT1-3} appeared in Ref.~\cite{Restrepo:2013aga}.

\subsection{Discussion}
\begin{figure}[t]
\centering
\begin{tabular}{ccccc}
\epsfig{file=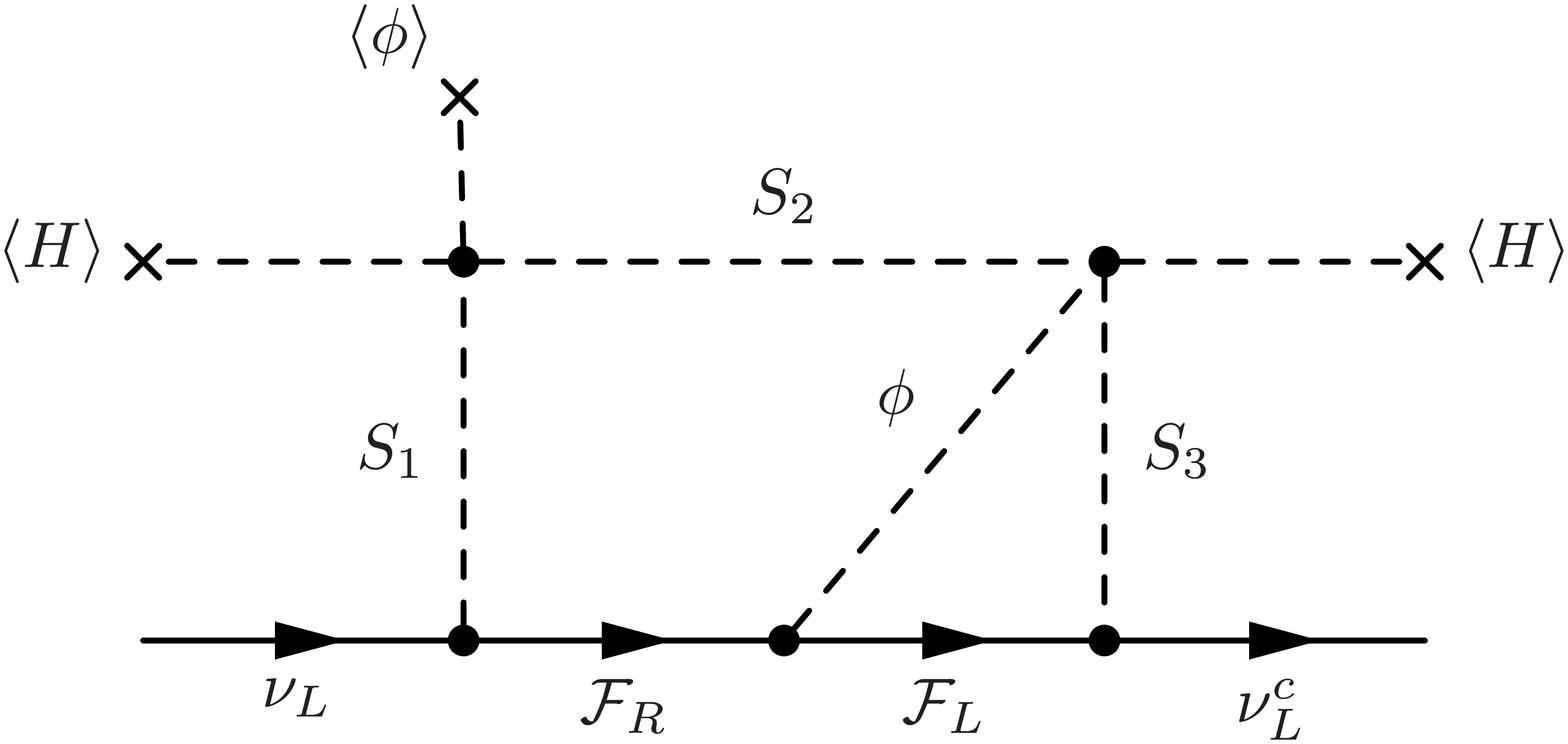,width=0.3\linewidth,clip=} & & %
\epsfig{file=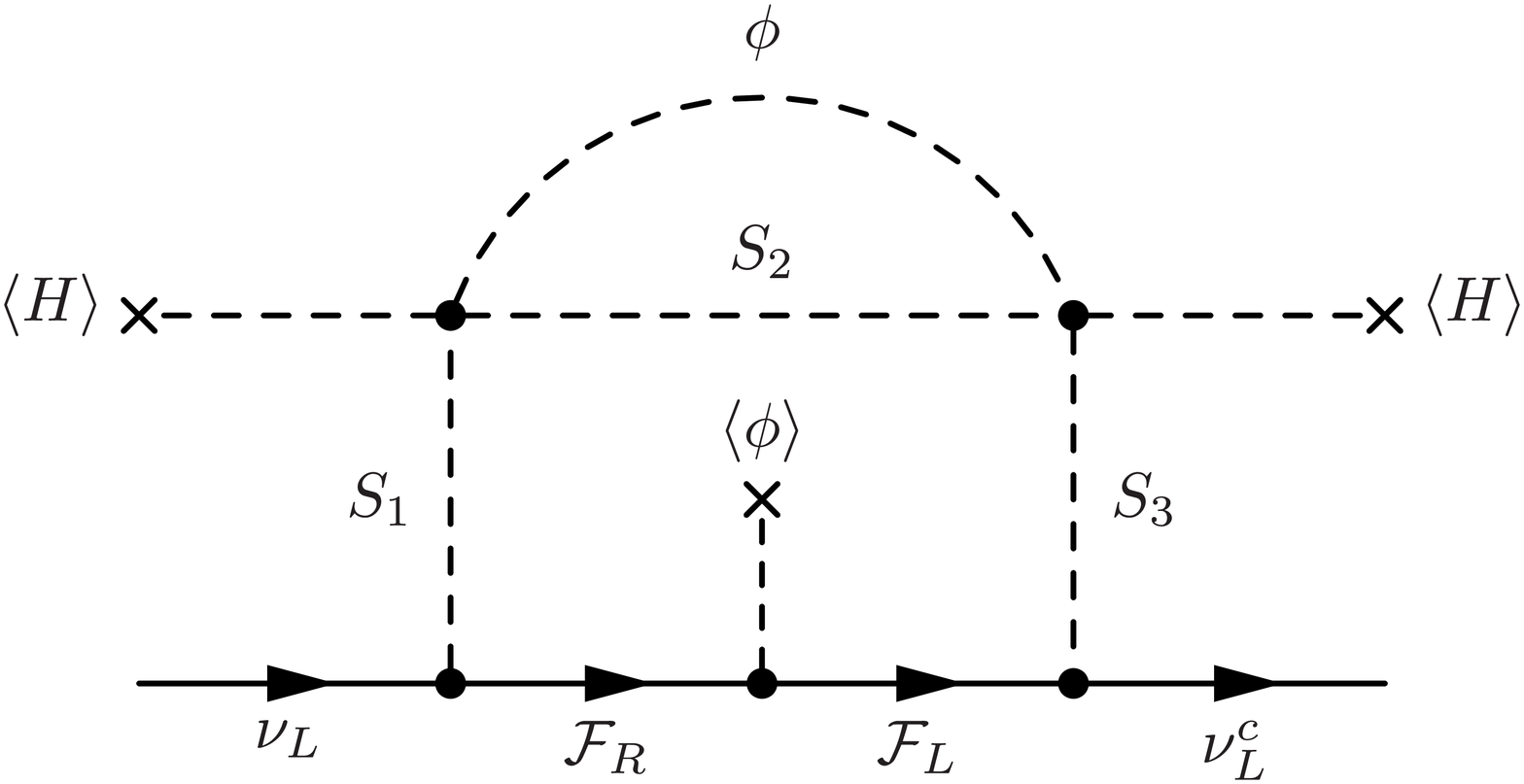,width=0.3\linewidth,clip=} & & %
\epsfig{file=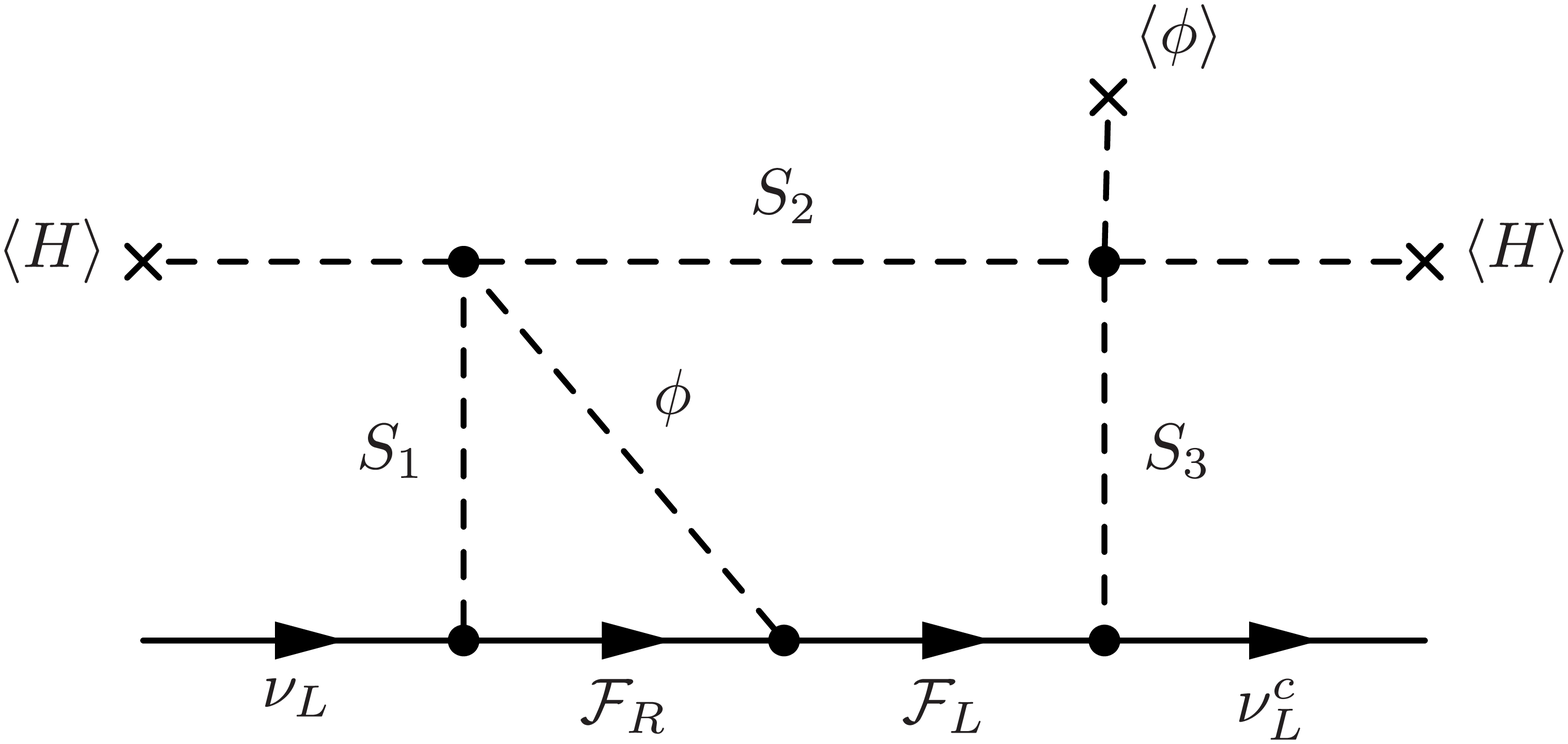,width=0.3\linewidth,clip=} \\
(a) & & (b) & & (c) \\
& & & &
\end{tabular}%
\caption{Two-loop diagrams obtained by closing scalar lines on the scale-invariant type T1-i diagram. Diagram $(a)/(c)$ is the right-/left-leaning two-loop diagram, diagram $(b)$ is the Robotman. }
\label{fig:T1-1two_loop}
\end{figure}

We observed earlier that the SI type T1-i topology gives a neutrino mass diagram with mass-dimension eight, different from the other topologies, whose diagrams have mass-dimension  six. This allows one to close pairs of external $\phi$ lines to form the two-loop diagrams in Figure~\ref{fig:T1-1two_loop}. These diagrams have mass-dimension six, which is less than the one-loop diagram, so it is not \emph{a~priori} evident that they are suppressed relative to the one-loop diagram (higher-loop diagrams with lower mass-dimension can be comparable to diagrams with higher mass-dimension, see~e.g.~Ref.~\cite{Kumericki:2012bh}). The $d=8$ one-loop diagram contains an additional factor of $\sim(\langle \phi\rangle/M)^2$, where $M$ denotes a generic mass for exotics in the loop. In a general theory, such factors could suppress the $d=8$ one-loop diagram relative to the $d=6$ two-loop diagram, such that the additional loop-suppression of the latter is overcome. However, in our SI models all mass scales are related to the scalar VEVS and one has $M\lesssim\langle\phi\rangle$, so the factor $\sim(\langle \phi\rangle/M)^2$ should not give a significant suppression. Thus, we naively expect the one-loop diagram to dominate the two-loop diagrams.\footnote{For completeness, we note that the two-loop diagram obtained from Figure~\ref{fig:SI_T1-1} by closing the external Higgs lines via the $|H|^4$ vertex still has mass-dimension $d=8$ and is thus suppressed by an additional loop factor relative to the one-loop $d=8$ diagram in Figure~\ref{fig:SI_T1-1}.} Nonetheless, strictly speaking the SI T1-i models generate neutrino mass by a combination of one-loop diagrams of mass-dimension eight and two-loop diagrams of mass-dimension six. Depending on taste, one may wish to use a labeling scheme that distinguishes the SI T1-i topology from the other one-loop topologies, as they are not ``pure" one-loop models. For our purposes we retain the labeling scheme of Ref.~\cite{Bonnet:2012kz} for ease of comparison, though we note the difference. 

More generally, promoting any $n$-loop non-SI neutrino mass diagram to a minimal SI implementation will also allow diagrams with $>n$ loops if the non-SI diagram contains more than one fermion mass insertion or cubic scalar coupling.  On the other hand, if one restricts attention to pure one-loop models, then only three minimal topologies count, namely  SI T3, T1-ii and T1-iii, giving fifteen distinct models with DM candidates. 

In a related matter, note that we did not include the SI implementation of the model in Ref.~\cite{Law:2013dya}, with real scalar triplet, in our lists. That model is similarly a hybrid one- and two-loop model, giving a $d=7$  one-loop diagram and a $d=5$ two-loop diagram. It appears that the SI implementation would amount to adding a real scalar triplet to the SI model T1-ii$(a)$ if the triplet is $Z_2$-even, while for a $Z_2$-odd triplet one arrives at the SI model T3$(d)$.

This concludes the classification of minimal SI one-loop diagrams. Inspection of the lists reveals that the simplest cases are the SI scotogenic model and the related triplet variant (see Table~\ref{table:general_SIMa}),  both of which require three beyond-SM multiplets (five, if one includes generation structure for $\f$). The SI T3 models with complex fermions and the SI T1-iii models both require four new multiplets, while the SI T1-i and T1-ii models require five new multiplets.

Before concluding, we note that, in general, one could ask whether additional distinct SI diagrams can be obtained by attaching $\phi$ VEV insertions to particle lines in the SI  T1 and T3 diagrams, or if other variations are possible. If one adds a single insertion of $\langle\phi\rangle$ to a scalar line, SI demands that the new vertex includes another new scalar, which can only be closed by forming a diagram with more than one loop. Alternatively, adding two $\langle\phi\rangle$ insertions at a single vertex requires that the other scalars at that vertex have the same quantum numbers, making the loop diagram essentially the same but with a higher mass-dimension (and number of fields, if new fields are added). Similarly, one can add two individual $\langle\phi\rangle$ insertions on a fermion line (two are needed to flip chirality twice), giving the same diagram but with higher mass-dimension. We found no other one-loop structures that correspond to effective operators with mass-dimension six. Thus, in summary, the minimal SI one-loop diagrams have a single singlet-VEV insertion, corresponding to the SI T3, T1-ii and T1-iii topologies, while the SI T1-i diagram has three insertions (equivalently, it gives two-loop diagrams with a single insertion).

\section{Conclusion\label{sec:conc}}

We have categorized the minimal irreducible SI one-loop topologies for neutrino mass and described the particle content for models that contain viable DM candidates. In all, we presented fifteen distinct SI models for one-loop neutrino mass with DM.  The models generically predict a new scalar $\phi$ that fulfills the dual roles of triggering electroweak symmetry  breaking and allowing lepton number symmetry violation to give radiative neutrino mass. This scalar also mixes with the SM Higgs, with important phenomenological consequences. The DM is part of a $Z_2$-odd sector, the content of which is model dependent. There are cases with singlet, doublet and triplet DM, and all predict new physics at the TeV scale. Models with triplet DM may require a degree of tuning, as the DM is typically $M_\dm=\mathrm{few}~$TeV, and all other $Z_2$-odd exotics must be heavier than this scale. However, even if one neglects models with triplets, multiple cases with singlet and/or doublet DM were found.

If one restricts attention to pure one-loop models, only three minimal irreducible SI one-loop topologies appear possible, namely the SI  T3,  T1-ii and  T1-iii topologies.  This differs from the non-SI case, where four distinct irreducible one-loop topologies are found~\cite{Bonnet:2012kz}.  Unlike the non-SI case, the SI implementation of the type T1-i topology has mass-dimension eight and generically allows for a two-loop neutrino mass diagram of lower mass-dimension, in addition to the SI T1-i diagram, producing a hybrid model. The T1-i  models also contain a subset of particles that generates the simpler SI T3 diagram.

As a check of our results, one can compare our list of models with irreducible SI one-loop topologies to the non-SI one-loop models with DM~\cite{Restrepo:2013aga}. Retaining the SI T1-i models, for comparison, our list remains considerably shorter than the corresponding non-SI result, where more than thirty models were found.\footnote{We neglect model T1-i-C in Ref.~\cite{Restrepo:2013aga}, with $\alpha=1$, as it fails to realize neutrino mass.}  There are important differences between the SI and non-SI cases; for example, with the SI type T1-iii topology, scalar DM is generally not viable, whereas the corresponding non-SI T1-iii models  admit scalar DM. However, such differences do not account for the discrepancy in the overall number of models; all such cases retain a viable DM candidate for both the SI and non-SI models, and thus appear on both lists. Instead, the discrepancy results from our neglect of type T1 models in which a type T3 diagram is also generated, as these  were considered as generalized versions of the T3 models.

Finally, we note that we categorized the minimal irreducible SI one-loop topologies for neutrino mass, giving explicit quantum numbers for exotics that include a viable DM candidate. However, our classification of the distinct SI one-loop diagrams  is not dependent on whether the loop diagram contains a DM candidate. If one is not considering a relationship between DM and neutrino mass, the fields in the loop diagram need not transform under a discrete symmetry and thus may include SM fields. In such cases, the diagrams in Figures~\ref{fig:general_SI_1loop},~\ref{fig:SI_T1-2} and~\ref{fig:SI_T1-3} would simply show the minimal  irreducible SI  one-loop topologies for neutrino mass,\footnote{Presumably the quantum numbers of the exotics would  be selected to prevent tree-level neutrino mass, though this is a model dependent issue.}  while Figure~\ref{fig:SI_T1-1} would give a further irreducible one-loop topology that permits a two-loop diagram with lower mass-dimension.

\section*{Acknowledgments\label{sec:ackn}}
AA is supported by the Algerian Ministry of Higher Education and
Scientific Research under the CNEPRU Project No D01720130042. KM is supported by the Australian Research Council.
\appendix


\end{document}